\documentclass[12pt]{JHEP3}
\usepackage{amsmath,epsfig}
\def\be{\begin{equation}}
\def\ee{\end{equation}}
\def\ba{\begin{eqnarray}}
\def\ea{\end{eqnarray}}

\def\beq{\begin{equation}}
\def\eeq{\end{equation}}
\def\beqa{\begin{eqnarray}}
\def\eeqa{\end{eqnarray}}

\def\preal{{\rm Re\,}}
\def\pim{{\rm Im\,}}

\def\yzero{\smash{\hbox{$y\kern-4pt\raise1pt\hbox{${}^\circ$}$}}}
\def\p{\partial}
\def\a{\alpha}
\def\b{\beta}
\def\g{\gamma}
\def\d{\delta}
\def\m{\mu}

\def\cV{{\cal V}}
\def\beq{\begin{equation}}
\def\eeq{\end{equation}}
\def\beqa{\begin{eqnarray}}
\def\eeqa{\end{eqnarray}}

\def\vt{\vartheta}

\def\-{\hphantom{-}}

\def\s2{\frac{1}{\sqrt2}}

\def\oh{\frac{1}{2}}
\def\beq{\begin{equation}}
\def\eeq{\end{equation}}
\def\beqa{\begin{eqnarray}}
\def\eeqa{\end{eqnarray}}

\def\Tr{{\rm Tr \,}}

\def\IF{\relax{\rm I\kern-.18em F}}
\def\II{\relax{\rm I\kern-.18em I}}
\def\IP{\relax{\rm I\kern-.18em P}}
\def\IC{\relax\hbox{\kern.25em$\inbar\kern-.3em{\rm C}$}}
\def\IR{\relax{\rm I\kern-.18em R}}

\def\cn{{\cal N}}

\def\Vol{{\rm Vol}}

\def\Dsl{\,\raise.15ex\hbox{/}\mkern-13.5mu D} %this one can be subscripted
\def\IZ{Z\kern-.4em  Z}

\def\fn{\footnote}
\def\idty{{\bf 1}}

\def\inte{{\bf Z}}

\def\cA{{\cal A}}

\def\ti{\tilde}

\def\w{\wedge}

\title{Wilson Line Inflation}

\author{ A. Avgoustidis, D. Cremades, F. Quevedo \\ 
{\it   Department of Applied Mathematics and Theoretical
Physics,\\ Wilberforce Road, Cambridge CB3 0WA, UK\\}
\email{A.Avgoustidis, D.Cremades, F.Quevedo@damtp.cam.ac.uk}} \abstract{
We present a general set-up for inflation in string theory where the
inflaton field corresponds to Wilson lines in compact space in the
presence of magnetic fluxes.  $T$-dualities
and limits on the value of the magnetic fluxes relate this system to the
standard D-brane inflation scenarios, such as brane-antibrane
inflation, D3/D7 brane
inflation and different configurations of branes at angles. This can
then be seen as a generalised approach to 
inflation from open string modes. 
Inflation ends when the Wilson lines achieve a critical
value and an open string mode becomes tachyonic. Then 
hybrid-like inflation, including its cosmic string remnants, is
realized in string theory beyond the brane 
annihilation picture.
Our formalism can be incorporated within
flux-induced moduli stabilisation mechanisms in type IIB
strings. Also, contrary to the standard D-brane separation, Wilson
lines can  be considered in heterotic string models. We provide explicit
examples to illustrate similarities and differences of 
our mechanism  to D-brane inflation. In particular we
present an example in which the $\eta$ problem present in brane
inflation models is absent in our case. We have examples with both
blue and red tilted spectral index and remnant  cosmic string tension 
$G\mu \lesssim 10^{-7}$.}

\preprint{DAMTP-2006-22\\  hep-th/0606031}

\begin{document}

\section{Introduction}
There are several classes of moduli fields in string
compactifications. Closed string moduli include the geometric and
dilaton fields, with the geometric ones separated into the
deformations of the  K\"ahler structure and deformations of the
complex structure for Calabi-Yau and related compactifications. Open string moduli include the location of branes
as well as Wilson lines which correspond to background values of gauge
fields, usually coming from the open string sector of the type I/II
string theory. In
heterotic string however,  Wilson lines are closed string modes.

Most of these fields have been proposed as inflatons in terms of explicit
mechanisms to realize cosmological inflation in string theory. There
are now concrete models for which the inflaton is a K\"ahler modulus
\cite{racetrack, conlon} and several scenarios in which the modulus is
a brane separation \cite{dvalitye,burgess, d3d7, rabadan, bcsq, dbi}. 
A general property of brane inflation models is that inflation
terminates by means of a tachyon condensation mechanism similar to
hybrid inflation. This open string tachyon itself has been also
proposed to be the inflaton field in other scenarios \cite{cqs}.
Each of these  scenarios has distinct physical implications which can be put to test
  by observations, regarding remnants of inflation such as cosmic
  strings, the value of the spectral index, etc.

Obtaining several realizations of inflation within string theory is a
major achievement of the past five years.
It would be highly desirable to have a general set-up that includes at
least general classes of scenarios and try to extract as much model
independent
 predictions. At the moment we are still at the stage of
exploring the different possible realizations of inflation within
string theory. For instance, there is not yet a concrete proposal for 
having the complex structure moduli as inflatons, something that could
be
desirable. Also Wilson lines have not been yet used as
inflatons in an explicit string construction\fn{Notice however that  the authors of \cite{berg} computed loop corrections to a variation of 
the KKLMMT model in terms of Wilson lines. Also, the possibility of using Wilson lines as inflatons was considered in \cite{creminelli}
in a five dimensional effective field theory setup.}.
The purpose of this paper is to explore this possibility.

Wilson lines correspond to background values of gauge fields that can
be turned on in spaces of non-trivial homotopy. They provide most of
the moduli in toroidal compactifications of the heterotic string 
parametrising the Narain lattices \cite{nsw}. Discrete and continuous
Wilson lines
 have also been used
to break the gauge symmetries in phenomenological attempts to
realistic models \cite{GSW, inq, aiq}. Furthermore, in the context of
D-branes, Wilson lines have been playing a crucial role to understand
dualities in type I/II strings \cite{polchi}. A $T$-duality transforms branes at
angles in such a way that the angles are mapped to magnetic fluxes and 
D-brane separations are mapped to Wilson lines in the dual model. Since
brane separations have been successfully used as inflatons, it is
natural to study the Wilson lines as inflatons. Furthermore, this 
formalism is the one that can naturally extend to the heterotic case
in which the D-brane separation has no direct analogue.

We present our results as follows. Section 2 is dedicated to introduce
the basic ideas and derive, in the case of toroidal
compactifications, the string amplitude 
 induced when magnetic fields and Wilson lines are turned on.
We also comment on the fact that all potentials used in the literature as starting points for 
brane/antibrane inflation, D3/D7 inflation and D-branes at angles can be obtained from the amplitude we consider
after performing $T$-dualities in different directions, as well as
taking limits on the values of the magnetic fields. 
This means that all toroidal D-brane inflationary configurations can be obtained starting only with D9 branes
with magnetic fluxes and Wilson lines in Type I string theory.
The next section is dedicated to explore the Wilson line potential
extracted from this amplitude
 for inflation and a concrete model of inflation is presented.
 Section 4 discusses the supergravity description of this set-up and
 compares with the KKLMMT approach to brane antibrane inflation in the
 context of moduli stabilization. We present our conclusions in
 section 5. We have an appendix 
 explaining in detail the duality between Wilson lines and brane
 separation as well as the structure of Wilson lines in general.
A further appendix discusses an explicit model in which the dilaton, complex
structure and the location of D7 branes are fixed by RR and NS-NS
fluxes but Wilson lines remain flat, leaving them as the right
candidates for the inflaton field.

\section{Wilson Lines and String Amplitudes}

The aim of this section is to show how the vacuum energy generated
in a supersymmetry breaking system of branes can have a dependence
on Wilson line degrees of freedom generated in the world-volume of
the branes. For the sake of clarity, let us stick to a simple
configuration that illustrates the general idea. Consider IIB string
theory compactified on a $T^6$, with a set of D7-branes filling the
four dimensional space-time and wrapping the same toroidal four
cycle in this $T^6$. The $T^6$ is
constructed as ${\bf R}^6/\Gamma$, with $\Gamma$ being the lattice
generated by six vectors $\{e_i\}$, $i=1,...,6$. One can use
affine parameters along these axes as $x^i,\in(-\pi,\pi)$, and we will 
set the metric so that the length of each lattice vector $e_i$ is $2\pi R_i$.  
To begin with we will consider the torus to be factorisable as a product of
square tori, but later on we will generalise this setup.  

In the absence of further orbifold/orientifold
projections, this system preserves 16 supercharges and thus it leads
to $\cn=4$ supersymmetry when reduced to four dimensions. The
existence of non-trivial 1-cycles in the $T^4$ wrapped by the
D-branes leads to the presence of Wilson line degrees of freedom,
four for each brane\fn{There are other scalar degrees of freedom in
the lower dimensional theory, apart from these Wilson lines and the
position of the branes in the $T^2$
transverse to their worldvolume, corresponding to strings going from
one brane to a different one. A vev for these scalar fields triggers
a rank-reducing bifundamental gauge symmetry breaking, and
is the field theory counterpart of the geometrical recombination of the system of
branes. The only place where these fields will play a role in the
following is in the symmetry breaking putting an end to inflation in
some of the models, as it will be explained below, and thus one can
consider their vevs to be set to zero during most of what follows.},
that show themselves as the lowest components of chiral multiplets
in the reduced theory. These scalar fields have flat potentials and
thus they can take an arbitrary vev consistently with supersymmetry,
but with the consequence that these vevs can lead to rank-preserving
adjoint gauge symmetry breaking, as explained in modern
textbooks (see e.g. \cite{polchi}).

This kind of flat potentials are known to be lifted in
supersymmetry-breaking situations. This is indeed the case here. As
we will see, the presence of some source of supersymmetry breaking,
such as a non-zero magnetic field in the world-volume of one of the
branes, is able to generate a potential for these previously
shift-symmetric Wilson lines degrees of freedom\fn{This shift-symmetry is of course related to the gauge invariance in the 
world-volume of the D7 and consequently the potential will break this gauge invariance. The point here is that the potential between the first brane 
(call it $a$) and the second one (named $b$) will depend on the difference 
of Wilson lines, $\lambda_a-\lambda_b$, but not on the sum of them. 
When $\lambda_a-\lambda_b$ reaches some particular value (given 
in (\ref{mass_tach})), a tachyon with charge $(+1a,-1b)$ develops, 
leading the system to a final configuration
in which the $U(1)$ gauge group with charge $Q_a-Q_b$ is broken. Since the potential is independent of $\lambda_a+\lambda_b$, the corresponding gauge symmetry associated
to the group with charge $Q_a+Q_b$ is preserved all along the process.}.

For simplicity, let us restrict ourselves to a system with just two
D7 branes, wrapping the same $T^4$ in the $T^6$. Let the coordinates
of the $T^6$ be given by $\{x^i\}$, $i=1,...,6$, and assume the
pair of D7-branes to be wrapping the second and third torus ($i=3,...,6$), being
point-like (and one on top of each other) in the first one. Consider
turning on a Wilson line along one of the directions of the $T^4$,
say $x^5$, in only one of the branes\fn{So that
$U_\gamma=e^{i\lambda}$. In the following we will use the term
Wilson line to denote both $U_\gamma$ and $A_\gamma=(\lambda/L)
d\sigma$, with $\sigma$ being the affine parameter parametrising the
curve $\gamma$, $L$ the length of $\g$ and $\lambda\in(-\pi,\pi)$.
Moreover, from the four dimensional point of view, $A_\gamma$
corresponds just to the zero mode of the field we will generically
denote as Wilson line degree of freedom. We hope this abuse of
language will not confuse the reader.} (the extension to a more
general Wilson line is straightforward):
\beqa A={\lambda \over 2\pi R_5}dx^5. \label{field}\eeqa
with $\lambda\in(-\pi,\pi)$. The presence of this Wilson line will
break the gauge group $U(2)\to U(1)$. As stressed, in the absence of
any source of supersymmetry breaking one has a flat potential for
$\lambda$. A way of lifting this flat potential is to add a source
of supersymmetry breaking, such as magnetic or B-field in the
world-volume of the brane. The addition of
magnetic field or the presence of B-field in the world-volume of the
brane leads to a FI term in the four dimensional theory
\beqa \xi\sim \int_\Sigma J\w(B+2\pi\a' F) \, , \eeqa 
where $\Sigma$ is the four-cycle wrapped by the D-brane, $J$ is the
K\"ahler form of the compactification space, $B$ is the pullback of
the ambient B-field on $\Sigma$ and $F$ is the magnetic field in the
world-volume of the D-brane. This FI term will generically break
supersymmetry, and the system will try to evolve towards its
restoration, generating non-trivial potentials for the fields
involved. 

\subsection{The interaction potential}
Consider a magnetic field in the world volume of one of the D7-branes of the form
\beqa
F={2\pi m\over A_2}dx^3\w dx^4
\eeqa
with $m\in \inte$ and $A_2$ the area of the corresponding 2-cycle.
In this situation, a Wilson line-dependent vacuum energy is
generated, that can be computed via the Coleman-Weinberg formula,
see \cite{cliff, polchi, arfaei}
\beqa\label{ampl}
\cA=2\times \cV_{d+1}\int {d^{d+1}k\over (2\pi)^{d+1}}\int_0^\infty {dt \over 2t} \Tr e^{-2\pi \a't(k^2+M^2)} ,
\eeqa
where $d$ stands for the number of non-compact directions. The extra factor of 2 comes as usual from interchanging the two ends
of the string. Note that the integral in the
momenta $k$ must only be computed for the {\it external} dimensions,
since for the internal ones the momentum is a quantised quantity. We
will then include the sum over the internal momentum in the trace of
the mass operator.

For generality, we will start from 
a more general configuration of two D$p$-branes, wrapping the same  
$T^{(p-3)}$, with magnetic flux turned on in a $T^2$ submanifold wrapped by  
one of them.  We allow for Wilson lines to be turned on along the $n$  
compact dimensions parallel to the branes in which there is no magnetic flux turned on, 
and we associate a brane separation degree
of freedom $x_i$ to each of the $9-p=4-n$ transverse directions. The mass operator for the twisted string states
lying between both branes is given by

\beqa \alpha' M_{ab}^2 =
\sum_{i=1}^{4-n}{\Delta_i^2\ti R_i^2\over 4\pi^2 \a'}+\sum_{i=1}^n
{P_i^2\a' \over 4\pi^2R_i^2}+ N_\nu + \nu ( \theta_{ab} - 1 )\ ,
\label{mass2} \eeqa
where the number operator $N_\nu$ can be found\fn{$\nu=0,\oh$ for the R and NS sector, respectively.} in \cite{fernando}, and we have  
called $R_i$, $i=1,...,n$ the compactification radii corresponding to the  
$n$ dimensions in which Wilson lines can be switched on, $\tilde R_j$,  
$j=1,...,4-n$ the radii corresponding to the transverse directions. The  
number $\theta$ is defined in analogy with 
the intersecting brane case (see appendix) as 
\beq
\tan\theta=\frac{(2\pi)^2 \a' m}{A_{\rm flux}},
\eeq
 with $A_{\rm flux}$ being the area of the torus and $m$ the number of quanta of magnetic flux,
and
\beqa
\Delta_i^2&=&(x_i+2\pi w_i)^2\\
P_i^2&=&(\lambda_i+2\pi m_i)^2
\eeqa
with $x_i,\lambda_i\in (-\pi,\pi)$. The amplitude (\ref{ampl}) can be 
written as\footnote{There is also a spacetime volume factor  
$\cV_4$ which does not appear in the final expression for the 
potential, so we have not included it in the amplitude.} 
\beqa\label{amplitude} 
\cA&=&\int_0^\infty{dt \over t}\left(\int{d^{4}k\over
(2\pi)^{4}} e^{-2\pi\a'tk^2}\right)
\sum_{m_i,w_i}e^{-{\Delta_i^2\ti R_i^2 t\over 2\pi \a'}}e^{-P_i^2\a't\over  
2\pi R_i^2} \Tr e^{-2\pi\a'(N_\nu+\nu(\vt_{ab}-1))}\nonumber\\
&=&{1\over (8\pi^2 \a')^2}\int_0^\infty {dt \over t^3} \sum_{m_i,w_i}\exp\left(-\sum_{i=1}^{4-n}{\Delta_i^2\ti R_i^2t\over 2\pi \a'}\right)\exp\left(-\sum_{i=1}^n{P_i^2\a't\over 2\pi R_i^2}\right)
Z(\theta,t)\nonumber\\
\,  
\eeqa
with\fn{We are using the definitions of \cite{polchi}.}
\beqa
Z(\theta,t)={\vartheta_{11}^4(i |\theta| t/2\pi,it)\over
i\vartheta_{11}(i|\theta| t/\pi,it)\, \eta^9(it)}. \nonumber\eeqa
The behaviour of this function $Z(\theta,t)$ for $\theta\neq 0$ in the large and small $t$ limit is
\beqa
Z(\theta,t)&\to& e^{|\theta| t}, \hspace{4.3cm} t\to \infty,\\
Z(\theta,t)&\to& 4t^3\sin^2\left({|\theta|\over 2}\right)\tan\left({|\theta|\over 2}\right),\hspace{.5cm} t\to 0.
\eeqa
Note that the mass of the lowest lying string state between both branes is given by
\beqa\label{mass_tach} 
M^2=\sum_{i=1}^{4-n}{x_i^2\ti R_i^2\over 4\pi^2 \a'^2}+\sum_{i=1}^n {\lambda_i^2\over 4\pi^2R_i^2}-{|\theta|\over 2\pi\a'},
\eeqa
and this implies that, 
provided that the infinite sums converge, we are left with an infinite  
series of converging integrals, whenever the mass is not tachyonic. The  
presence of the exponentials allows us to substitute $Z(\theta,t)$ by  
its low $t$ limit and we get (assuming in the following $\theta>0$)
\beqa
\cA={ \sin^2({\theta/ 2})\tan({\theta/ 2})\over(8\pi^2)^2\a'^n}{\prod_{i=1}^n R_i\over \prod_{i=1}^{4-n}\ti R_i}\prod_{i=1}^{4-n}\exp(iw_ix_i)\prod_{i=1}^n
\exp(im_i\lambda_i)\nonumber\\ \times\int_0^\infty {dt\over t^2} \sum_{m_i,w_i}\exp\left(-\sum_{i=1}^{4-n}{\pi \a'\over 2\ti R_i^2t}w_i^2-\sum_{i=1}^n{\pi R_i^2\over 2\a't}m_i^2\right)\label{integral} , 
\eeqa
where we have used the Poisson resummation formula
\beqa
\sum_{n\in\inte}e^{-\pi a n^2+\pi b n}={1\over \sqrt{a}}\sum_{n\in\inte}e^{-{\pi\over a}\left(n+i{b\over 2}\right)^2}.
\eeqa
The first term (with $m_i=w_i=0$ $\forall i$) is divergent. This  
divergence is due to the sum over all images in a compact space, and, as  
explained in \cite{rabadan_zamora}, it is an unphysical divergence\fn{The  
function we obtain after the regularisation is (apart from numerical factors) 
a solution to the Green's equation in the torus,
given by $\nabla^2 G(x)=\d(x)-1/V$, with $V$ the volume of the torus. The sum over images corresponds to an infinite sum over solutions of the non-compact Green's equation
and thus one cannot expect it to give the correct answer; this is the origin 
of this non-physical inifinity.}.
After discarding this infinity, we get
\beqa
\cA_{\rm reg}={ \sin^2({\theta/ 2})\tan({\theta/ 2})\over(8\pi^2)^2\a'^n}{\prod_{i=1}^n R_i\over \prod_{i=1}^{4-n}\ti R_i}\sum_{(m_i,w_i)\neq (0,...,0)}{\prod_{i=1}^{4-n}\exp(iw_ix_i)\prod_{i=1}^n
\exp(im_i\lambda_i)\over \sum_{i=1}^{4-n}{\pi \a'\over 2\ti R_i^2}w_i^2+\sum_{i=1}^n{\pi R_i^2\over 2\a'}m_i^2}.\nonumber\\
\label{full_potential}
\eeqa
The interaction term is then
\beqa
V_{\rm int}=-\cA_{\rm reg}.
\eeqa
Since we want to apply this to the D7 case, we set $n=2$.

The potential (\ref{full_potential}) corresponds to the solution of 
Poisson's equation with a $\delta$-function source, that is the 
Green's function, on the torus.  Due to the symmetry  
of the torus, the Green's function is not rotationally symmetric  
at distances of order the compactification scale away from the source
\cite{rabadan} .  We now concentrate on the behaviour of
the potential near the core $(x_i=\lambda_i=0)$, where the rotational  
symmetry is restored.  The infinite sum becomes
\beqa
\sum_{(m_i,w_i)\neq (0,...,0)}{\prod_{i=1}^{4-n}\exp(iw_ix_i)\prod_{i=1}^n
\exp(im_i\lambda_i)\over \sum_{i=1}^{4-n}{\pi \a'\over 2\ti R_i^2}w_i^2+\sum_{i=1}^n{\pi R_i^2\over 2\a'}m_i^2}\simeq {2 \pi^3 V\over ||X,\Lambda||^2}
\eeqa
with
\beqa
||X,\Lambda||^2=\sum_{i=1}^2{x_i^2\ti R_i^2\over \a'}+\sum_{j=1}^2{\lambda_j^2\a'\over R_j^2}\equiv X^2+\Lambda^2 ,
\eeqa
and
\beqa
V={4\over \pi^2}{\prod_{i=1}^2\ti R_i\over \prod_{i=1}^2 R_i}  , 
\eeqa
so that the final amplitude reads
\beqa\label{fin_ampl} 
{\cal A}_{\rm reg}={ \sin^2({\theta/ 2})\tan({\theta/ 2})\over 8\pi^3\a'^2  
||X,\Lambda||^2} . 
\eeqa
This is, as expected, of the same form as the potential of the first  
reference of \cite{rabadan}.  As we will see in the next section, the  
phenomenology in our setup can be quite different from that of \cite{rabadan}.  The inflaton in our case will be (up to a normalization) $\Lambda$ rather than 
$||X,\Lambda||$ and the fact that $\Lambda\propto 1/R_i$,  
$||X||\propto \ti R_i$ implies that, for generic choices of brane 
separations $x_i$, we have $||X||\gg \Lambda$.  Then, expanding around 
$\Lambda/||X||=0$ yields a quadratic, rather than Coulombic, potential.  
In the limit of small brane separations such that $||X||\ll \Lambda$ the 
potential is indeed Coulombic.    

The computations we have done here are valid for square tori. In more general situations one might want to consider the effect of adding a real part to the complex structure
of the torus. If the complex coordinates of a $T^{2n}$ are given by $dz^i=dx^i+\tau^i dy^i$, then the Green's function is given by
\beqa
G(z_i)=-{1\over \prod_{i=1}^n (2\pi R_i)^2\pim \tau_i}\sum_{(n_i,m_i)\neq(0,0)}{\prod_{i=1}^n\exp\left({2\pi i \pim(n_i\bar z_i \tau_i+m_iz_i)\over \pim \tau_i}\right)
\over \sum_{i=1}^n {|m_i-n_i\tau_i|^2\over R_i^2\pim\tau_i^2}},
\eeqa
and computing the corresponding inflationary potential is straightforward along the lines sketched here.

\subsection{The Constant Term}\label{constant}

The full inflationary potential will consist of a constant part, which gives the vacuum energy density responsible for inflation, 
plus an interaction, slowly decreasing, term. 
The interaction term was computed in the previous section. To compute the 
constant term, we will assume that the energy
of the system after inflation has ended is of order the cosmological 
constant today (hence near zero); this means that all possible 
contributions to the vacuum energy included in
the system, such as other branes, orientifolds and fluxes, sum up to 
zero\fn{Note that here we are assuming a local configuration, since 
we are mainly 
interested in the dynamics associated to the Wilson lines. The dynamics 
associated to the full system can be very complicated and its analysis 
is beyond the 
scope of this work. This, in particular, means that we are not 
explicitly cancelling RR tadpoles, an issue that should be addressed 
in a complete model.}.

The vacuum contribution can be found by considering the energy difference  
between the chosen configuration and the minimum energy configuration with  
the same charges.  Here, the configuration of interest involves two  
D7-branes with magnetic flux\footnote{Wilson lines do not contribute  
to the vacuum part of the potential at this level but, as we will see 
in section \ref{sugra}, moduli stabilization effects can induce such 
a dependence.} in a $T^2$ wrapped by one of them, but we can work in 
the T-dual picture of branes at angles and then T-dualise back. Consider 
for example two D$1$-branes wrapping $(n_1,m_1)$ and $(n_2,m_2)$ cycles 
respectively in a $T^2$.  If $R_1, R_2$ are the radii of the torus then 
the energy of the configuration reads  
\be
  E=TR_1\left[\sqrt{n_1^2+m_1^2 R_2^2/R_1^2}+\sqrt{n_2^2+m_2^2 R_2^2/R_1^2}
  \right] 
\ee
with $T$ the corresponding brane tension. The minimum energy state corresponds to a reconnected configuration of 
one brane wrapping a $(n_1+n_2, m_1+m_2)$ cycle 
\be 
 E_{\rm min}=TR_1\sqrt{(n_1+n_2)^2+(m_1+m_2)^2 R_2^2/R_1^2} . 
\ee 
In the case of interest $n_1=n_2=1, m_1=0, m_2=m$ so that 
\beqa
 \Delta E&=&TR_1 \left(1+\sqrt{1+m^2 R_2^2/R_1^2}-2\sqrt{1+m^2 R_2^2/(4R_1^2)} 
 \right)\nonumber\\
 &\simeq& \frac{1}{4}TR_1\frac{m^2 R_2^2}{R_1^2} . 
\eeqa 
Now, T-dualising in the $R_2$ direction, we have $R_2\rightarrow 
\frac{\a'}{R_2}$ so $\Delta E$ becomes 
\be
 \Delta E \simeq \frac{1}{4}TR_1\frac{m^2 \a'^2}{R_1^2 R_2^2} .
\ee
Similarly, for D7-branes wrapping a $T^4$ of volume $V_4$ as well  
as four spacetime dimensions, and with a magnetic flux switched on  
in a $T^2$ submanifold with radii $R_1$ and $R_2$, we have a vacuum 
energy density 
\be
 \Delta E \simeq \frac{1}{4}T_7 V_4 \frac{m^2 \a'^2}{R_1^2 R_2^2}\equiv
 \frac{1}{4}T_7 V_4\frac{(2\pi)^4 m^2 \a'^2}{V_{\rm flux}^2}  ,   
\ee
where $V_{\rm flux}=(2\pi)^2 R_1 R_2$ is the volume of the $T^2$ in which 
the flux is turned on. $T_7$ is the D7-brane tension 
%in the Einstein frame, 
given (for $p=7$) by
\beqa
T_p={(2\pi)^{-p}\a'^{-{p+1\over 2}}\over g_s}.
\label{tension}
\eeqa
This corresponds to a contribution to the potential 
\be
 V_0\simeq {(2\pi)^{-7}\a'^{-4}V_4\over 4 g_s}\left({(2\pi)^2 \a' m\over  
 V_{\rm flux}}\right)^2 \equiv {(2\pi)^{-7}\a'^{-4}V_4 \over 4 g_s} 
 \tan^2\theta , 
\ee
where we parametrised the magnetic flux in terms of an angle $\theta$,  
the same angle appearing in the interaction potential, cf. (\ref{fin_ampl}).    

The full potential (in terms of $\lambda$) is the sum of the vacuum and 
interaction terms 
\beqa\label{potential} 
V(\lambda)=V_0+V_{\rm int}(\lambda).
\eeqa

\subsection{Unifying Models}\label{models}
Though straightforward, it is interesting to comment on the fact 
that most of the open string inflationary potentials
(without the contributions of fluxes) may be regarded as coming from a
 common origin, namely Type I string theory with magnetic
 fluxes\footnote{We 
 refer  the reader to the end of appendix \ref{ap_duality}
for the conventions relevant to this section.}.

Consider Type I string theory on a $T^6$ with possibly magnetic flux in 
some of the cycles. Apart from the D9 branes included in Type I theory, 
all other possible branes can be seen as different configurations of magnetic flux on the world volume of these D9-branes. It follows that one can get any toroidal inflationary model
previously considered in the literature just starting from different configurations of D9 branes with magnetic flux and Wilson lines in Type I theory.
\begin{itemize}
\item{One of the possible ways of getting brane-antibrane inflation from this set-up is putting infinite magnetic flux 
in one of the 2-cycles wrapped by some brane and minus infinite magnetic flux in the same 2-cycle in some
other brane. The inflaton will be a Wilson line (or a combination of them) along some 1-cycle.}
\item{The model of inflation from branes at angles \cite{rabadan} can be obtained starting from Type I just considering a small amount of magnetic flux in some two-cycle
wrapped by a D9 and
making three T-dualities along three one-cycles, one of them belonging to the two cycle that supports the flux. The variation
of this model considered in the last reference of \cite{rabadan} can be obtained considering magnetic flux in two different two-cycles. The inflaton(s) in both cases will be
a (combination of) Wilson lines along one-cycles not supporting magnetic flux.}
\item{The model of D3-D7 inflation \cite{d3d7} in a toroidal set-up can be obtained starting with magnetic flux along two 2-cycles in Type I and then performing six T-dualities.
Again, the inflaton will be a Wilson line along some appropriate 1-cycle transverse to the two-cycles where magnetic flux has been turned on.}
\item{One can try to generalise this picture considering more general compactification spaces that either have one-cycles or alternatively considering lower ($p<9$) dimensional branes wrapping
some cycles having non-trivial one-cycles on them, to be able to turn on Wilson lines on them. Magnetic flux along some two-cycle will break supersymmetry generating a non-trivial potential for
the Wilson line.}
\end{itemize}

\section{Inflation from Wilson Lines}\label{sec_WL}

In this section we study the inflationary properties of the above 
potential, assuming that the volumes of all two-cycles in the compactification manifold have been  
fixed. In principle this is a possibility to consider, since there is no 
obstacle preventing this kind of mechanisms to
occur in string theory, as commented in \cite{kklmmt}. However, most known
mechanisms to fix volumes proceed along the lines of superpotential 
stabilisation of K\"ahler moduli, and this situation will 
be analysed in section \ref{sugra}. Along the present section we will 
simply assume that the volumes of the two-cycles have been stabilised 
by some unspecified mechanism. 

We consider a specific set-up, in the spirit of the previous section,  
in which two D7-branes wrap a $T^4$ submanifold of a compactification 
manifold $T^6$.  We write $T^6$ as a product of three 2-tori with radii 
$(R_4,R_5), (R_6,R_7)$ and $(R_8, R_9)$ respectively, and arrange the 
branes to be pointlike in the first torus, with separations $x_4, x_5$.  
In the world-volume of one of the branes, we turn on magnetic flux  
(parametrised by the angle $\theta$) in the second torus and a Wilson  
line along the 8-direction of the third torus.  With this notation 
the potential becomes
\be\label{pot_inf} 
 V(\lambda) = {R_6R_7R_8R_9 \over  8\pi^3 \a'^4 g_s} \frac{\tan^2(\theta)}{4} 
 -{ \sin^2({\theta/ 2})\tan({\theta/ 2})\over 8\pi^3\a'^2 ||X,\Lambda||^2} 
\ee        
with 
\be
 ||X,\Lambda||^2={x_4^2 R_4^2\over \a'}+{x_5^2 R_5^2\over \a'}+ 
 {\lambda^2\a'\over R_8^2}\equiv X^2+\Lambda^2 . 
\ee

\subsection{Normalization and Inflationary Parameters}

We now move to analysing the inflationary properties of the potential  
(\ref{pot_inf}).  We assume that we have fixed the brane separations  
$x_4, x_5$ by fluxes (see appendix \ref{app_model}) but not the Wilson  
line $\lambda$, which will play the role of the inflaton.  In order  
to compute the slow-roll parameters we first need to identify the  
canonically normalised field.  The kinetic term for the Wilson line  
comes precisely from the first term in the expansion of the DBI action,  
namely the Maxwell action.  We have
\beqa
S=\int d^4x \sqrt{-g} \left(-{1\over 4g_{YM,4}^2} F_{\mu i}F^{\mu i}+...\right)=- \int d^4x  \sqrt{-g} \left({1\over 2g_{YM,4}^2} \left(\p_\mu A_i\right)^2+...\right)\nonumber\\
\eeqa
with $g_{YM,4}$ the (four dimensional) Yang-Mills gauge coupling constant, where we have used the fact that the Wilson lines have a
constant profile in the internal dimensions. The relation between $g_{YM,4}$ and other parameters in the model is
\beqa
{1\over g_{YM,4}^2}=T_p(2\pi \a')^2 \Vol_{p-3}
\eeqa
with $\Vol_{p-3}$ the volume of the internal dimensions wrapped by the D-brane and $T_p$ the D-brane tension:
\beqa
T_p={(2\pi)^{-p}\a'^{-{p+1\over 2}}\over g_s}.
\eeqa
The relation between $A_i$ and the numbers $\lambda_i$ is given by \cite{polchi}
\beqa
A_i={\lambda_i\over 2\pi R_i}.
\eeqa
%As we have seen, however, the potential depends on the $\lambda_i$ only in the specific combination
%\beqa
%\phi^2\equiv \sum_{i=1}^n {\lambda_i^2\over R_i^2}.
%\eeqa
In this simplest case we have turned on only one Wilson line degree of 
freedom $\lambda$, along the 8-direction of the third torus corresponding 
to radius $R_8$.  Then $\lambda$, the (non-canonically normalised) inflaton,
will appear in the Lagrangian with a kinetic term
\beqa
S_{{\rm k},\phi}=-\int d^4x \sqrt{-g} \oh \left( T_p \a'^2 \Vol_{p-3}\right) 
\left(\p_\mu{\lambda\over R_8}\right)^2  . 
\eeqa
We have not pulled the $1/R_8$ out of the derivative for reasons that will  
become apparent afterwards.  We can then define a canonically normalised  
field $\psi\equiv K^{1/2} \lambda$, with
\beqa
K\equiv { T_7 \a'^2 V_{4}/R_i^2} \, , 
\eeqa
where we have already substituted $p=7$. If we assume that we can fix the volumes of the different two-cycles, then we already have all the ingredients to compute the inflationary parameters.  These are given by
\beqa
\epsilon\ =\ {M_p^2\over 2}\left({V'\over V}\right)^2,\qquad \eta\ =\
M_p^2\,  {V''\over V}
\eeqa
where the prime means differentiation with respect to the canonically normalised field. It is convenient to recall the expression of the Planck mass in terms of stringy quantities,
for unwarped compactifications. It is given by
\beqa\label{Mpl_stringy}
{M_p^2\over 2}={(2\pi)^{-7}\over g_s^2\a'^4}\Vol_6  , 
\eeqa
where $\Vol_6$ stands for the total six-dimensional volume.  Then, with 
$\psi=K^{1/2} \lambda$, we get
\beqa
\epsilon&=&{M_p^2\over {2K}} ~{1\over V^2}\left({dV\over d\lambda}\right)^2 
={(2\pi)^2R_4R_5R_8^2\over {g_s \a'^2}} ~{1\over V^2}\left({dV\over d\lambda} 
\right)^2\label{eps_inf}\\
\eta&=&{M_p^2\over K} ~{1\over V^2}{d^2V\over d\lambda^2}={2(2\pi)^2R_4R_5R_8^2 
\over {g_s \a'^2}} ~{1\over V}{d^2V\over d\lambda^2}\label{eta_inf} . 
\eeqa

\subsection{Inflation Phenomenology}

Before exploring whether the slow-roll conditions are satisfied in  
regions of the parameter space, let us pause and consider the qualitative 
features of the potential.  First, we observe that  $||X||$ scales with  
radius and $\Lambda$ with inverse radius, so for generic brane separations 
and similar compactification scales, we have $\Lambda\ll ||X||$.        
Expanding $||X,\Lambda||^{-2}$ around $\Lambda/||X||=0$ we find a  
quadratic term in $\lambda$ with positive mass.  Inflation can
then be realised by starting with $\lambda>0$ and rolling towards the
$\lambda=0$ minimum.  We can further arrange (see section \ref{tachyon}) 
that a tachyonic instability appears when the field reaches a critical 
value $\lambda_{\rm crit}$.  Thus the inflationary phase comes to an end  
in a hybrid inflation fashion.  

We now explore the conditions for slow-roll inflation.  Using the potential
(\ref{pot_inf}) and the approximation $V\simeq V_0$ in the 
denominator\footnote{In (\ref{pot_inf}) the presence of compactification  
radii, $g_s$, $||X,\Lambda||$ and $\theta$, all lead to a suppression 
of $V_{\rm int}$ relative to $V_0$}, equation (\ref{eps_inf}) reads
\beq\label{epsilon1}
\epsilon=4(8\pi)^2 g_s \frac{R_4 R_5\,\a'^4}{R_6^2 R_7^2 R_9^2 R_8^4}
\frac{\sin^4(\theta/2)\tan^2(\theta/2)}{\tan^4\theta}
\frac{\lambda^2}{||X,\Lambda||^8} . 
\eeq
Similarly, equation (\ref{eta_inf}) for $\eta$ becomes
\beq\label{eta1}
\eta = (8\pi)^2 \frac{R_4 R_5 \, \alpha'}{R_6 R_7 R_8 R_9}
\frac{\sin^2(\theta/2)\tan(\theta/2)}{\tan^2\theta}
\frac{1}{||X,\Lambda||^4} \left(1-\frac{4\alpha'  \lambda^2}{R_8^2
||X,\Lambda||^2} \right) .  
\eeq
The slow-roll conditions $\epsilon\ll 1, \eta\ll 1$ can then be satisfied
by choosing appropriately the relevant compactification volumes and, more
importantly, by tuning the angle $\theta$ to be sufficiently small.

Note however that the choice of compactification volumes and angle  
needs to be consistent with the COBE normalization constraint.  Indeed,  
if the inflaton $\lambda$ is responsible for generating the cosmological
density perturbations, then the observed CMB anisotropies as measured
by the COBE satellite, impose the normalization
\beq
\delta_{\rm H}\simeq \frac{1}{5\sqrt{3}\pi M_p^2} \sqrt{\frac{V}{\epsilon}} 
=1.91\times 10^{-5},
\eeq
or equivalently \cite{LiddleLyth} 
\beq
\left(\frac{V}{\epsilon}\right)^{1/4}=0.027 M_p .
\eeq
Approximating $V\simeq V_0$ during slow-roll and using equation
(\ref{Mpl_stringy}) we obtain a constraint for the average 
compactification radius
\beq\label{R_COBE}
\bar R \sim \left(\frac{g_s^3 \tan^2\theta}{\epsilon}\right)^{1/8}
\sqrt{\alpha'}.
\eeq
Then, equation (\ref{Mpl_stringy}) gives for the string scale
\beq\label{Ms_COBE}
\sqrt{\alpha^\prime} M_p \sim \left(\frac{g_s}{\epsilon^3}\right)^{1/8}
(\tan\theta)^{3/4}.
\eeq
For the example we will consider below with $\theta\sim 0.1$ and  
$g_s\sim 0.1$, we will find $\epsilon \sim 10^{-11}$ resulting in a  
(low) GUT string scale, as is typically the case in brane inflation models.   

We highlight that the constraint (\ref{R_COBE}) involves the angle $\theta$
as well as the compactification scale(s), so one does not have unlimited
freedom in choosing these parameters to satisfy slow-roll.  As we will now 
see, requiring that inflation ends in a hybrid-type exit further constrains  
the possible parameter choices.

\subsection{Graceful Exit}\label{tachyon}

An attractive feature of the model is the possession of a natural mechanism 
to exit inflation, via a tachyonic instability.  The mass of the lowest 
string state stretching between the branes is (see equation  
(\ref{mass_tach})) 
\beqa\label{mass_tach_mod}
m^2=\sum_{i=4}^{5}{x_i^2 R_i^2\over 4\pi^2 \a'^2}+{\lambda^2\over  
4\pi^2R_8^2}-{|\theta|\over 2\pi\a'} .
\eeqa
We observe that, if $R_4, R_5$ and $\theta$ are chosen appropriately, 
then the mass squared of this mode becomes negative as the inflaton 
rolls towards $\lambda=0$.  Beyond this critical point, an 
instability appears and the field quickly rolls down in the tachyonic 
direction, violating slow-roll and thus ending the inflationary 
phase.  This provides an elegant realization of hybrid inflation  
beyond the brane annihilation picture.  Furthermore, since the tachyon  
is charged under the gauge groups on the branes, its condensation 
allows the formation of cosmic strings via the Kibble mechanism  
\cite{burgess,strings}.       

It seems that the additional tuning of parameters required to guarantee  
a hybrid inflation type exit could be inconsistent with the COBE  
normalization or violate the slow-roll conditions, preventing 
inflation.  Indeed, for a tachyon to develop, the first and last terms  
in equation (\ref{mass_tach_mod}) need to be comparable and this seems to  
require small compactification radii and large angles, which is the opposite 
of what is needed for slow-roll (equations (\ref{epsilon1}-\ref{eta1})).      
However, equation (\ref{mass_tach_mod}) only involves $R_4, R_5$ and $R_8$ 
so the rest of the radii can still be much larger to guarantee that 
slow-roll is satisfied.  Furthermore, the first term in (\ref{mass_tach_mod}) 
can be made small by putting the branes closer together (adjusting $x_4, x_5$ 
small) rather than reducing the compactification radii $R_4, R_5$.  As a 
result, fine tuning can be achieved at the order of one part in 1000 to 
guarantee slow-roll, while satisfying the COBE normalisation and ensuring 
that tachyon condensation takes place after an appropriate number of  
e-folds.  In the next section we present an explicit example of such a 
model.

\subsection{An Explicit Example} 

We now consider a specific choice of parameters, which gives rise to 
slow-roll and tachyon condensation, while satisfying the COBE constraint.   
Most of the fine tuning comes from requiring that the tachyon condensation  
mechanism kicks in at the right value of $\lambda$.  Indeed, $\lambda$ 
ranges from $-\pi$ to $\pi$ and to make sure that (\ref{mass_tach_mod}) 
changes sign for some critical value $\lambda_{\rm crit}$ in this range  
(in fact we need $\lambda_{\rm crit}<1$ to get enough e-folds) requires  
that the first and last terms in equation (\ref{mass_tach_mod}) are equal  
in magnitude to an accuracy of three significant figures.  This corresponds  
to fine tuning of a few parts in a thousand.  In our example we will choose 
$R_4=R_5=R_8=5\sqrt{\a'}$ and magnetic flux such that $\theta=0.4$.  By 
choosing a configuration in which the branes are only separated in the 
$4$-direction, setting $x_4=0.317$ and $x_5=0$, we obtain 
$\lambda_{\rm crit}=0.161$.  Then, as long as we are able to satisfy the 
slow-roll conditions, there is enough room for inflation between 
$\lambda\approx 1$ and $\lambda_{\rm crit}$.  

In order to satisfy slow-roll we can choose the rest of the  
compactification radii $R_6, R_7$ and $R_9$ to be large.  As already  
mentioned however, we cannot make the slow-roll parameters arbitrarily small 
because the compactification scale is constrained by the COBE normalization.  
Fortunately, equation (\ref{R_COBE}) allows for large enough radii to give 
rise to enough inflation.  Small string coupling $g_s$ leads to a  
smaller $\epsilon$ but does not affect $\eta$.  Choosing for example  
$R_6=R_7=R_9=20\sqrt{\a'}$ and $g_s=0.1$ gives $\epsilon(\lambda\sim 1)  
\approx 10^{-11}$, $\eta(\lambda\sim 1)\approx 10^{-3}$.  Then, 
starting inflation at $\lambda=0.5$, for example, gives rise to 
$N\simeq 400$ e-folds of inflation until the  
inflaton rolls down to $\lambda_{\rm crit}$, where tachyon condensation  
kicks in.  Cosmological scales exit the horizon around 60 e-folds before the 
end of inflation (corresponding to $\lambda\simeq 0.19$) at which point the 
slow-roll parameters are: 
\beqa\label{epsilon_model} 
&\epsilon\simeq 4.6 \times 10^{-13} \\ 
&\eta\simeq 2.8 \times 10^{-3} .  
\eeqa    
The scalar spectral index is then $n_s=1-6\epsilon+2\eta \simeq 1.006$.   
Note that although such a Harrison-Zel'dovich spectrum is disfavoured,  
it is still consistent with the combined WMAP3+SDSS data \cite{KKolbMRiot}.   
Also, if cosmic strings are indeed produced in the end of inflation as 
described in the previous section, then one should include the contribution
of the string network when comparing to the CMB, in which case a flat 
inflationary spectrum plus strings is consistent, if not favoured, 
by the data~\cite{BHKU,BGMS}. 
Due to the smallness of $\epsilon$, the model predicts no significant  
running of the spectral index and no gravitational waves.    

There are many other consistent choices of parameters with similar 
predictions.  We can for example start with a different angle $\theta$ and  
modify $R_4, R_5$ accordingly, or select the brane separations $x_4, x_5$  
differently, to get the tachyonic instability at an appropriate value 
$\lambda_{\rm crit}$.  We can also choose a smaller $R_8$ to make the  
effect of the second term in equation (\ref{mass_tach_mod}) more important,  
thus relaxing somewhat the fine tuning of brane separations.  The rest 
of the radii are chosen so as to give rise to slow-roll, and simultaneously 
satisfy the COBE normalisation.  Exploring the parameter space one can easily  
arrange $\epsilon\lesssim 10^{-10}$,  but the $\eta$ parameter is  
typically much bigger $\eta\gtrsim 10^{-3}$.  Thus, a  
Harrison-Zel'dovich/slightly blue spectrum of scalar perturbations and 
no significant running or gravitational waves are robust predictions of 
the model in this limit.  

The above picture is valid for relatively large angles, when the brane 
separations needed to arrange a satisfactory hybrid inflation model are 
such that $||X||>\Lambda$.  For small enough angles, successful tachyonic  
condensation can only happen for $||X||<\Lambda$.  In this case the   
potential is not of the $1+\phi^2/\m^2$ form but instead goes like 
$\phi^{-2}$, in close analogy to reference \cite{rabadan}.  This leads  
to very different predictions than the above situation, in particular it  
gives rise to negative $\eta$ and hence a slightly red scalar spectrum,  
in better agreement with WMAP3.  For example, having $R_4=R_5=5\sqrt{\a'}$,  
$R_8=2\sqrt{\a'}$ and $R_6=R_7=R_9=40\sqrt{\a'}$ with $\theta=0.001$ and  
$g_s\simeq 0.1$ yields $\epsilon(\lambda\simeq 1) \approx 10^{-12}$,  
$\eta(\lambda\simeq 1)\approx -3\times10^{-3}$.  This gives rise to  
$N\approx 100$ e-folds of inflation from $\lambda\simeq 0.7$ to  
$\lambda_{\rm crit}\simeq 0.16$.  The slow roll parameters 60 e-folds  
before inflation are $\epsilon\simeq 4\times 10^{-11}$ and $\eta\simeq  
-0.012$, corresponding to a spectral index $n_s\simeq 0.976$.  The  
smallness of $\theta$ follows from the requirement\footnote{For  
large angles, the third term in equation (\ref{mass_tach_mod})  
cannot be balanced by the Wilson line term.} that the tachyonic  
regime appears at an acceptable value $\lambda_{\rm crit}<1$.    

One comment is in order here. Apart from extra contributions to the $\eta$ parameter that will
appear when one considers dynamical moduli stabilisation mechanism 
(see section \ref{sugra}), there is one extra contribution that has
to be considered even in the present case. The interacting part of 
the inflationary potential is not given by (\ref{pot_inf}), as 
already emphasized in
the text, but rather by (\ref{full_potential}). Whereas both 
potentials are numerically close in the region of interest, it 
turns out that they are not equivalent
from the point of view of the parameter $\eta$. As showed in \cite{kklmmt} 
the first of these potentials (\ref{pot_inf}), roughly speaking, follows 
as a solution of a Green's equation of the form
$\nabla^2 V_1(x)=\d(x)$, which implies that the $\eta$ parameter far 
from the source at $x=0$ can be small, especially if there are 
more parameters to play with, like the
fluxes in our case. However, a potential of the form (\ref{full_potential}) 
will rather satisfy an equation of the form $\nabla^2 V_2(x)=\d(x)-1/V$, 
giving typically a contribution to
$\eta$ of order 1. This is certainly true case for brane-antibrane 
potentials. A detailed analysis \cite{tye} showed that this is not the 
case for branes at angles, an analysis that 
applies also here. The easiest way of seeing this is that, in our case, the interaction potential fulfills $\nabla^2_x V_{int}(y)=-\sin^2(\theta/2)\tan(\theta/2)/V$ whereas 
the constant part goes like $\tan^2(\theta)$, indicating that even when considering the complete (compact) potential the $\eta$ parameter will be suppressed. 
%It is interesting to note that the contribution they find to $\eta$ is of the same order of magnitude as ours, so even in the worst case scenario where one must add
%both contributions we still get a phenomenological model of inflation, at least under the assumption of volumes stabilisation considered here. 
A full (necessary numerical) 
analysis of this issue is beyond the scope of this paper.

One may wonder how the above results would be affected by considering more 
complicated models, as for example having more than one Wilson lines.  This  
situation is interesting since one of the Wilson lines can be 
used as the inflaton, while the other could play the role of a curvaton
field, responsible for the generation of perturbations.  Also, as a two-field 
model, it could allow for significant non-gaussianity, but the investigation  
of this would involve the study of non-linear perturbations (see for example 
\cite{RigVanShel}).  We leave this study for a further publication.  One could 
also consider more realistic compactifications, or embedding a similar model 
in heterotic theory.  

\section{Supergravity Description and Moduli Stabilisation}\label{sugra}

The previous sections have been written in the spirit of the first
articles on brane inflation. Even though we discussed a model in the
appendix in which the dilaton, complex structure moduli and D7 brane
moduli
are fixed, we have assumed that all K\"ahler  moduli
have been fixed by some unknown mechanism and only concentrated  on the dynamics
of the inflaton field. This is a strong assumption. Fortunately there
has been much progress in fixing all geometric moduli by means of
RR and NS fluxes combined with non-perturbative effects for the
K\"ahler moduli. The prime
example is that of \cite{kklt}. Since  \cite{kklmmt}, the standard
approaches towards inflation now incorporate the dynamics of these
moduli fixing that plays an important role in obtaining  inflation. 

\subsection{Effective Action Generalities}

In effective supergravity theories after Calabi-Yau compactifications
we know that the D3 brane/antibrane system can be described in terms
of an effective field theory. This has been discussed in detail
starting from the work of \cite{kklmmt}. In this case the anti D-brane
is fixed at a location inside the Calabi-Yau, namely the tip of a Klebanov-Strassler
throat in a deformed conifold geometry, and
the D3 brane position is parametrised in terms of a scalar field
$\varphi$. The supergravity theory includes the 
geometric moduli fields and $\varphi$. Within the KKLT scenario
\cite{kklt} of moduli
stabilization, the effective field theory can be described in terms of
only the K\"ahler moduli and $\varphi$ after fluxes fix the dilaton and
complex structure moduli.

In the simplest one-modulus case, the K\"ahler potential takes the form 
\beq\label{kahler}
K=-3\ \log\left(T+T^*-\varphi^*\varphi \right)
\eeq
and the superpotential takes the KKLT form
\beq\label{super}
W\ =\ W_0\ + \ Ae^{-aT}
\eeq
With $W_0$ a flux superpotential, taken constant after complex
structure and dilaton stabilisation.
To the supersymmetric Lagrangian constructed out of $K$ and $W$ we
have to add the supersymmetry breaking interaction terms describing
the tension of the anti D3 brane and the Coulomb interaction between
the branes. This system has been analysed in some detail in the past
few years in which $\varphi$ can be the inflaton field
\cite{kklmmt,bcsq,tye}. 

In this
framework, moduli stabilization is fully considered when analyzing
the prospects for 
inflation\fn{For related work regarding moduli stabilisation in D3/D7 inflation see \cite{davis}.}. With this set-up it has been found that it is possible
to get inflation as long as some fine tuning of the parameters is
performed. The fine-tuning is required because of the standard 
$\eta$ problem
of $F$-term inflation. A conformally coupled scalar, such as $\varphi$
induces a contribution of order one to the slow-roll parameter $\eta$ and
therefore it needs to be compensated by fine tuning the parameters of
the model to obtain small $\eta$. More explicitly, for brane inflation
the potential is of the form:

\beq
V\ =\ V_0\left(\cal{V}\right)\ +\ V_{int}\left(\varphi, \cal{V}\right)
\eeq
where $\varphi$ is the candidate inflaton and ${\cal{V}}$ the volume 
modulus. For fixed volume, the first term $V_0>0$ is a constant that dominates the
potential and gives rise to almost de Sitter expansion, while $V_{int}$ 
is subdominant but by its dependence on $\varphi$ 
provides the slow-roll conditions. The $\eta$ problem appears in
these string models because the K\"ahler modulus that is fixed by the
KKLT mechanism is not just
the volume $\cal{V}$ but a combination of $\cal{V}$ and
$\varphi$. This then induces a $\varphi$ dependence in $V_0$ that
 gives rise to $\eta\sim{\cal{O}}(1)$.

A fine tuning is required to have terms say in $V_{int}$ that cancel
the order ${\cal{O}}(1)$ value of $\eta$ to one part in at least
$100$.
In general this fine tuning has been approached in several ways:
\begin{enumerate}
\item{}
Taking into account the full potential as a function of $T$ and
$\varphi$. The parameters $W_0, A, a$ as well as the warp factor of the
metric defining the throat can be fixed to find a point when both $V'$
and $V''$ vanish (where the primes are derivatives with respect to the
brane position field $\varphi$.). Perturbing around this point there is a
region in which both $\eta$ and $\epsilon$ satisfy the slow-roll
conditions and give rise to at least $60$ {\it e}-folds of inflation.
In \cite{bcsq} it was found that this fine tuning is of order
$1/1000$, worse by one order of magnitude than the expected
$1/100$. But more general considerations in the second article of
reference \cite{bcsq} reduced this
tuning. In both cases a multi-brane configuration was needed to
accommodate both, the scale of  inflation ($\sim 10^{15}$ GeV)
 and the standard model scale $1$ TeV.

\item{}
In \cite{d3d7} it is argued that inflation could appear naturally if there is a
shift symmetry for the field $\varphi$.

\item{} 
Non-perturbative corrections to the superpotential may depend on
$\varphi$ as first proposed in \cite{kklmmt}. This means that the
parameter $A$ can be $\varphi$-dependent. Explicit string calculations 
for simple models have been performed in \cite{berg} where this 
$\varphi$ dependence was found, providing an explicit way to fine 
tune.  

\item{}
Different configurations were proposed. In particular, having two
throats with anti D3 branes at the tip of each throat (as in second
article of \cite{shamit})
provided a way to tune the potential such as to cancel the order one
contribution to $\eta$ by the competition between the two anti branes
to attract the D3 brane.

\item{}
In \cite{dbi}, the full DBI action was used to obtain `fast-roll'
inflation for the brane/anti-brane system:
\beq
S_{DBI}\ = \ \int d^4x~
a^3(t)\left[V(\varphi)\sqrt{1-{\dot\varphi}^2/V(\varphi)} \ + \ U(\varphi)\right]
\eeq
With $T(\varphi)$ the space-dependent D3-brane tension. Its functional
form comes from the warp-factor dependence of the metric and
identifying the Calabi-Yau coordinate $r$ with the D3 brane location. The
function 
$U(\varphi)$ includes the mass term induced from the conformal coupling
as well as the interactions.

\end{enumerate}

We may wonder if similar approaches can be used for Wilson lines. It
so happens that the K\"ahler potential for Wilson lines in Calabi-Yau
orientifold compactifications of type IIB string theory, has the
same dependence on Wilson lines $\lambda$ as for the field $\varphi$
above \cite{louis}. Therefore we can use the K\"ahler potential above 
by substituting $\varphi\leftrightarrow \lambda$.
Since the superpotential is the same as for KKLT and, as we have seen
in the previous sections,  the interactions
have a similar dependence on the Wilson lines as for the Coulomb-like
interaction among branes, we conclude that the physics is very much
the same in both cases. An advantage of Wilson lines is that they
provide more parameters to be tuned, for example the value of the
fluxes of the magnetic fields. 

Furthermore, in \cite{berg} the corrections to the non-perturbative
superpotential
above were found actually for the T-dual model
in which the matter fields appearing in $W$ are
precisely the Wilson lines. This dependence on Wilson lines permits
the tuning to be made to reduce the value of $\eta$ as suggested
originally in \cite{kklmmt}. This applies directly to our case also.

  It is clear that  a similar dependence on the Wilson lines
appears as in
the original DBI action. It would be interesting to investigate how a
`potential' for the Wilson lines can be generated from the warp
factor.
But knowing the relationship between magnetised D7 branes and D3
branes, we expect a similar behaviour. Again, magnetised D7 branes will
offer more parameters to play in order to compare with observations.

\subsection{Wilson Line Inflation and Moduli Stabilisation}

Let us now describe the model of the previous chapter in terms of the
effective supergravity theory. As mentioned before, 
the known method to fix the K\"ahler parameters is the KKLT
set-up which uses a combination of fluxes and non-perturbative
superpotentials to fix the K\"ahler moduli. We can incorporate this to
our
type IIB toroidal model. Therefore we can have a proper treatment of
inflation in which there is a potential for all the fields and we
follow the  evolution of the candidate inflaton field.

Previous experience \cite{kklmmt,bcsq} shows that it is not possible
to just assume that this mechanism is at work and just consider
constant values of the K\"ahler moduli.
As mentioned in the previous subsection, the 
main fact to be taken into account is that 
what is fixed by this mechanism is not the volumes but a combination of the
volumes and the inflaton 
field. This is the source of the $\eta$ problem discussed in
\cite{kklmmt}. We see now how this problem is evaded in our example.

 In a factorisable
toroidal Type IIB set-up the supergravity fields are
\beqa
S&\equiv&-i\tau=e^{-\phi}+iC+ \oh \sum_a\sum_{i=1}^3|\zeta_a^{i,7i}|^2  \\
T_i&\equiv& {\oh e^{-\phi} A_jA_k}-i\int C_4\wedge \omega_i+\oh\sum_a\sum_{j,k=1}^3d_{ijk}|\varphi_a^{j,7k}|^2  \, \quad \, i \neq j \neq k\\
U_i&\equiv& \tau_i
\eeqa
with $A_i$ the area of the $i^{th}$ torus. $d_{ijk}=1$ if $i \neq j \neq k$ and zero otherwise. The $\zeta_a^{i,7i}$ are the supergravity fields corresponding to the transverse position of a $7_i$
brane, whereas the $\varphi_a^{j,7k}$ are the ones corresponding to
Wilson lines \footnote{Here we are following the standard notation (see
  for instance \cite{imr}) in which  $a$ labels the field and the
  index $j$ refers to the complex dimension  that has the Wilson line whereas $k$
  refers to the direction transverse to the D7 brane that hosts the
  Wilson line.}. The K\"ahler potential is given by
\beqa
K&=&-\log\left(S+S^*-\sum_a\sum_{i=1}^3|\zeta_a^{i,7i}|^2\right)-\sum_{i=1}^3\log\left(T_i+T_i^*-\sum_a\sum_{j,k=1}^3d_{ijk}|\varphi_a^{j,7k}|^2\right)\nonumber\\
&&-\sum_{i=1}^3\log(U_i+U_i^*)
\eeqa
The fact that the complex structure K\"ahler potential decouples is not completely clear but it is so at first order \cite{jockers}. Let us assume this form in what follows.
The outcome is that what we are fixing if we are applying some superpotential moduli fixing method are not the volumes but
\beqa
\preal T_i={\oh e^{-\phi} A_jA_k}+\oh\sum_a\sum_{j,k=1}^3d_{ijk}|\varphi_a^{j,7k}|^2.
\eeqa
In our particular case, let us say that the D7 is pointlike in the first torus $(k=1)$ and has flux in the second torus. Let us assume that we have been able to fix all the
$\preal T_i$. We have fixed then
\beqa
\preal T_1&=&\oh e^{-\phi}A_2A_3\\
\preal T_2&=& \oh e^{-\phi}A_1A_3+\oh |\varphi|^2\\
\preal T_3&=&\oh e^{-\phi}A_1A_2
\eeqa
This basically means that the values of all areas are given in terms of $|\varphi|^2$ as
\beqa
A_1&=&\sqrt{{2 e^\phi \preal T_3\over \preal T_1}\left(\preal T_2-\oh|\varphi|^2\right)}\\
A_2&=&\sqrt{{2 e^\phi \preal T_1 \preal T_3\over \preal T_2-\oh|\varphi|^2}}\\
A_3&=&\sqrt{{2 e^\phi \preal T_1\over \preal T_3}\left(\preal T_2-\oh|\varphi|^2\right)}
\eeqa
Assuming $|\varphi|^2 \ll 2(T_2+T_2^*)$, we obtain for the canonically 
normalised inflaton
\beqa
\psi=M_{p}{\varphi\over \sqrt{T_2+T_2^*}}.
\eeqa

Now, in order to see if there is an $\eta$ problem in this model, we have
to compute the value of $V_0$ in terms of the moduli $T_i$ and the
candidate inflaton field $\psi$.

%Since $|V_0|\gg V_{\rm int}$, we have $V(\lambda)\simeq V_0(\lambda)$ and we get
To express $V_0$ in the 4D Einstein frame we have to perform the standard 
Weyl transformation.  This is achieved by the metric rescaling:
\beq
g_{\mu\nu}\rightarrow \Omega g_{\mu\nu}
\eeq
with $\Omega= g_s^2/(A_1A_2A_3)$. Therefore the constant term in the
potential
will be the product of three contributions: the D7 brane tension 
$T_7V_4/g_s$ (with $V_4=A_3A_2$ the volume of the cycle wrapped 
by the D7 brane), 
the magnetic fluxes proportional to $m^2/A_2^2$ and $\Omega^2$ from
the Weyl rescaling. This gives 
\beqa 
V_0(\lambda)\, =\, T_7A_3A_2{\a'^2 g_s^4m^2\over 4 A_1^2 A_2^4
  A_3^2}={(2\pi)^{-3}m^2\a'^{-2}g_s^3\over 4 A_1^2 A_2^3 A_3} \,; 
\eeqa
expressed in terms of the fixed K\"ahler moduli this becomes
\beq V_0\ = \ {(2\pi)^{-3}m^2\a'^{-2}\over 32 \preal T_3^2 \preal T_1}\,,
%\preal T_3}(\preal T_2-\oh |\varphi|^2)\nonumber\\
%&=&{(2\pi)^{-3}m^2\a'^{-2}\over g_s}{\preal T_2\over \preal T_3}
%(1-{|\psi|^2\over M_{p}^2}), 
\eeq
which {\it does not} depend on $\varphi$. We can then see that there is
no contribution to a mass term for the inflaton field in $V_0$ that
was the source of the $\eta$ problem in \cite{kklmmt}.
Essentially, the model of section 3 is unaffected
by moduli stabilisation in this case, and we can just extract the same
conclusions regarding slow-roll parameters, density fluctuations and
the spectral index.

\section{Conclusions}

We have presented a  general set-up for inflation in string theory
with open string modes as inflatons. Wilson lines can play the
role of inflaton fields in a way similar to brane separations.
In fact Wilson lines are T-dual to brane separations and therefore
it is expected that they play a similar role. The physics of both
systems is the same but in different regimes (large against small
volume).
 Therefore if we are only interested on effective field theory
 descriptions then we would need
only the large internal volume potential on each formalism and they
are certainly not equivalent.

It is interesting that starting with a particular configuration of 
D-branes, fluxes and Wilson lines, all known proposals for open
string inflation can be included by T-duality and limits on the value
of the magnetic fluxes. Furthermore it is reassuring to know that the
end of inflation, with corresponding reheating and topological
remnants, such as cosmic strings, can go by the standard
tachyon condensation mechanism.

The explicit models presented in section 3 provide
 examples of more possibilities than
previously considered, such as the inclusion of the location of the D7
branes and the number of  parameters involved, like 
magnetic fluxes, that can 
parametrise a fully realistic treatment of inflation.
It is worth pointing out that both blue and red tilted values of the spectral index
can be obtained and that the string scale tends to be of order the GUT
 scale leading to the remnant cosmic string tension to be of order 
$G\mu \lesssim 10^{-7}$.

For a discussion with moduli fixing a la KKLT, our formalism is in
general very
similar to the brane/antibrane case, including the need to confront
the $\eta$ problem. But again there are more parameters that
can be varied to contrast with experimental signatures such as the
number of efoldings, the spectral index and the COBE normalized
$\delta\rho/\rho$. Remarkably, we found that the $\varphi$
dependence in $V_0$ does cancel for the example of section 3
after including moduli stabilisation, as discussed in section 4,
and therefore there is no $\eta$ problem, so the phenomenological 
results of section 3 hold after moduli stabilisation.
Notice that even though there is no $\eta$ problem in these models,
inflation is certainly not generic, as we needed to have the moduli, 
both geometric and D7 positions, as well as the magnetic fluxes, 
in particular ranges
in order to satisfy the requirements of slow-roll, tachyon
condensation and COBE normalisation. 
Still, it is encouraging that the fine-tuning required by the $\eta$
problem generic in D-brane inflation models is not present. 
It would be interesting to understand how general this cancellation
is. For this it would be worth exploring more general Calabi-Yau
compactifications
with four cycles having non-trivial two-cycles carrying either
magnetic fluxes or Wilson lines.

Notice that without a moduli fixing 
scheme, branes at angles provided a more flexible approach to
inflation than the brane/antibrane system. However in the extension to 
include moduli fixing, only the brane/antibrane system has been
considered so far. %This is in part due to the fact that 
%the intersecting brane constructions have been mostly done in 
%type IIA string theory whereas the discussions on moduli stabilisation
%have been mostly done for the type IIB theory.
From the discussion above we can see that this omission
is corrected here but  in terms of the dual version with Wilson
lines, instead of brane separation,  as inflatons and magnetic fluxes 
representing the brane angles.

There are further open questions that may be worth exploring following
our results. An explicit calculation on the heterotic string would be 
interesting, to have a concrete realization of inflation in the heterotic
case similar to brane inflation in type II. For this nonsupersymmetric
compactifications with Wilson lines would naturally provide both, the 
constant  and interaction terms as in the case we discussed in
section 3. 
Also, it would be interesting to check how the same cancellation we found to obviate the $\eta$ problem holds in T-dual configurations
in terms of branes at angles. 
Finally, as stressed in the appendix, Wilson lines will be present whenever the homotopy of the submanifold wrapped by a given D-brane is not trivial, so that
our mechanism is not restricted to the case of the torus.
There are known examples of compactification manifolds richer than tori, 
which allow for cycles with non-trivial 1-homology. It would be 
interesting to see how our mechanism
applies for these kind of manifolds. In particular, a complete 
discussion of our mechanism
including warped geometries would be desirable.

\section*{Acknowledgements}
We acknowledge useful conversations on the subject 
of this paper with S. Abdussalam, M. Berkooz, C.P. Burgess, 
P. G. C\'amara, J. Conlon, M.P. Garc\'{\i}a del Moral,  
F. Marchesano, R. Rabad\'an, E.P.S. Shellard, A. Sinha, K. Suruliz, 
and A. Uranga. D.C. thanks the Weizmann Institute for Science 
and the University of Zaragoza for hospitality during the completion of this  
work. A.A. is funded by the Cambridge European Trust and the Cambridge Newton Trust. The work of D.C. is supported by the University of Cambridge. F.Q. is  
partially funded by PPARC and a Royal Society Wolfson award.

\appendix

\section{General Aspects about Wilson Lines}

\subsection{Continuous and Discrete Wilson lines}
Given a gauge field $A$ defined over a manifold $M$, a Wilson line along a given closed path $\gamma=\p C$ is defined as
\beqa
U_\gamma=P \ \exp \oint_\gamma A.
\eeqa
In the Abelian case, the Wilson line is a gauge invariant quantity. We will restrict to this case henceforth. If $\gamma$ is a contractible loop, then
$U=P \exp \int_C F$ and $U=1$ whenever $F=0$. The interesting case comes when $\gamma$ is not contractible, since then one can have $F=0$ and $U\neq 1$ in a gauge invariant
way.

In principle, Wilson lines are associated to the first homotopy group of the manifold $\pi_1(M)$ \cite{GSW}. For each element of this homotopy group we have a non-contractible
1-cycle $\gamma$ and thus we can associate a Wilson line $U_\gamma$ to that cycle. However, most of the elements $U_\gamma$ are not independent and it is more useful to classify the
Wilson lines in terms of homology. The Wilson lines in a given manifold are classified\fn{We thank F. Marchesano for a clear explanation of this point.}
by group homomorphisms $H_1(M,\inte)\to U(1)$ \cite{joyce}. In general \cite{nakahara}
\beqa
H_1(M,\inte)\cong \underbrace{\inte \oplus ... \oplus \inte}_{b_1(M)}\oplus \inte_{k_1} \oplus ... \oplus \inte_{k_p}
\eeqa
for some $p$, where $b_1(M)=\dim ~H_1(M,{\bf R})$ is the first Betti number of $M$. The reason to use the homology group under the integers instead that under the real numbers in
the classification is that the torsion 1-cycles (associated to discrete Wilson lines) are invisible under $H_1(M,{\bf R})$. These discrete Wilson lines correspond to cycles that are
non trivial when going around them a given number $n-1$ of times but become trivial after going around $n$ times. All along this paper we will concentrate in continuous Wilson lines,
that are elements of $H_1(M,{\bf R})$.

Upon compactification of string theory in a given manifold $M$ in
the presence of D-branes, one will have continuous Wilson line degrees of freedom whenever $b_1(\Sigma)>0$, where $\Sigma$ is some $p-3$ cycle wrapped by a D$p$-brane (for
phenomenological reasons we will always consider D$p$ branes to be filling the four dimensional space-time and wrapping some $p-3$-cycle in the compact space $M$).
This is, then, not a generic situation in a CY, where $h^{0,1}=h^{1,0}=0$. It is possible,
however, to find one-cycles inside higher dimensional cycles in a CY, see \cite{mirror} for examples. Moreover, there are other spaces yielding $\cn=2$ supergravity in
$D=4$ upon compactification, like $SU(3)$ structure manifolds, where one can have $b_1(M)\neq 0$. Thus, Wilson lines are not generic in string theory
but they are common enough for not to be considered as a pathology of the torus.

\subsection{On the Duality Between Angled and Magnetised Branes}\label{ap_duality}
Consider Type II string theory compactified on a $T^2$, and a pair of intersecting branes making
angle $\theta$ in the directions $X_1$, $X_2$ that parametrise the torus. Without loss of generality, consider
the first of these D-branes (call it brane $a$) to be parallel to the direction $X_1$.
The boundary conditions for the string at $\sigma=0$ (corresponding to the brane $a$) and $\sigma=\pi$ (brane $b$) are
\beqa
\p_\tau X_1=\p_\sigma X_2&=&0, \, \quad \ \sigma=0,\nonumber\\
\p_\sigma(\cos \theta ~X_1 + \sin \theta~  X_2) &=&0\\
\p_\tau(-\sin \theta~ X_1 + \cos \theta~  X_2) &=&0, \, \quad \ \sigma=\pi.\label{angles}
\eeqa
These boundary conditions can be trivially obtained starting from a standard Neumann-Dirichlet boundary
condition in brane $b$ and performing a rotation of angle $\theta$.

On the other hand, consider the boundary conditions for a string lying between a D-brane filling a torus (brane $a$) , and another one (brane $b$)
filling the same torus but with constant magnetic field turned on in its
world-volume. The boundary conditions can be read from the worldsheet action (see e.g. \cite{Johnson}) and read

\beqa
\p_\tau X_1=\p_\sigma X_2&=&0, \, \quad \ \sigma=0,\nonumber\\
\p_\sigma X_1 + {\cal F}~ \p_\tau X_2 &=&0\nonumber\\
-{\cal F}~ \p_\tau X_1 + \p_\sigma X_2 &=&0,  \, \quad \ \sigma=\pi. \label{magnetic}
\eeqa
with ${\cal F}=2\pi\a'F_{12}$. We see then, comparing (\ref{angles}) with (\ref{magnetic}) that these descriptions are T-dual along the $X_2$ direction 
provided that\fn{Remember that T-duality along $X_i$ exchanges the action of $\p_\sigma$ and
$\p_\tau$ on $X_i$.}
\beqa
\tan \theta=2\pi\a'F_{12}.
\eeqa
In practice we will have more involved configurations that the one described above, namely IIA configurations with even dimensional branes intersecting at points in the compact space
(and filling the four dimensional space-time for obvious phenomenological reasons) or IIB constructions with odd-dimensional branes having non-trivial gauge bundles
in their world-volume. The easiest of those constructions is given by $T^6$ compactifications or orbifolds/orientifolds thereof. In the IIA side, the setup consists of a group of D6-branes,
each one wrapping an element of the 3-homology of the internal manifold; in the case of a $T^6=\prod_iT^2_i$ each brane is completely characterised by 7 integers: six of them come in pairs as
$\{(n^i,m^i)\}$, each one of them meaning the number of times the brane wraps each one of the directions of $T^2_i$ (say $n^i$ times wrapping the $x^i$ direction and
$m^i$ times wrapping the $y^i$ direction). The other one, named $N$,  
corresponds to the number of branes that are on top of each other and  
gives the rank of the gauge group living in the world volume of the brane.  
We obtain the IIB description of this setup T-dualising three times along  
the direction $x^i$. The mapping of a given $\{(n^i,m^i)\}$ configuration  
is as follows:
\begin{itemize}
\item{$N$ D6 branes specified by the vector $(1,0)(1,0)(1,0)$ in the IIA picture is mapped to $N$ D3 branes in the IIB picture. }
\item{$N$ D6 branes given by $(n,m)_i(1,0)_j(1,0)_k$, $m\neq 0$, $n$ and $m$ coprime, are mapped to a stack $Nm$ D5 branes filling the $i^{th}$ torus and point-like in the other two tori.
$n\neq 0$ implies the presence of units magnetic field in the D5 in the $i^{th}$ torus that break the rank of the gauge group from $Nm$ down to $m$. This magnetic field
is given by
\beqa
F={\pi n\over m}\idty_{N\times m}dx^i\w dy^i= {\pi i\over \pim \tau^i}{n\over m}\idty_{N\times m} \ dz^i \w d\bar{z}^{\bar i}
\eeqa
where $dz^i=dx^i+\tau^i dy^i$.}
\item{$N$ D6 branes characterised by a vector $(1,0)_i(n^j,m^j)_j(n^k,m^k)_k$, $m^j,m^k\neq 0$, $n^i$ and $m^i$ coprime for a given $i$, are mapped to a stack of
$Nm^jm^k$ D7 branes that are point-like in the $i^{th}$ torus and whose magnetic field is given by
\beqa
F=\sum_{a=j,k}{\pi i\over \pim \tau^a}{n^a\over m^im^j}\idty_{N\times m^j m^k} \ dz^a \w d\bar{z}^{\bar a}.
\eeqa
The rank of the gauge group is reduced from $Nm^jm^k$ down to $N$ because of the monopole background.}
\item{Finally, a stack of $N$ D6 branes specified by a vector $(n^1,m^1)(n^2,m^2)(n^3,m^3)$ with all $m^i\neq 0$ and $n^i$ and $m^i$ coprime for a given $i$
is mapped to a stack of $Nm^1m^2m^3$ D9 branes whose magnetic field is
\beqa
F=\sum_{a=1,2,3}{\pi i\over \pim \tau^a}{n^a\over m^1m^2m^3}\idty_{N\times m^1 m^2 m^3} \ dz^a \w d\bar{z}^{\bar a}.
\eeqa
Again, the rank of the gauge group is reduced from $Nm^1m^2m^3$.}
\end{itemize}

It is well known \cite{polchi} that under T-duality the
adjoint fields corresponding to the position of a brane in the direction under which the T-duality is performed are mapped to Wilson line fields living in the world-volume of the brane,
via the explicit mapping
\beqa
{\lambda\over R}\stackrel{\rm T}{\longleftrightarrow} {y\over \a'},
\eeqa
with $\lambda$ being the Wilson line, $y$ the position of the brane, and $2\pi R$ the length of the 1-cycle upon which we make the T-duality.

We must emphasize that we have performed the T-duality from Type IIA along the three $x^i$ axes. Alternatively, one can perform any other combination of dualities arriving to an 
equivalent situation with branes at angles and magnetised branes. One that is particularly interesting is starting from the Type IIA setup with branes at angles referred above and perform
three T-dualities along the $y^i$-axes. One reaches in this way Type I theory (or an orbifold thereof). Again one can have all kinds of D$p$-branes with $p$ odd and fluxes. The dictionary to
go from the Type IIB language quoted in this appendix and the Type I language\fn{This is actually an imprecision we make in order to differentiate between both images, since both of them
are Type IIB and are related by 6 T-dualities.} that is used in section \ref{models} is as follows\fn{One must stress that in order for this to be a real duality the complex
structure parameters $\tau^i$ must be imaginary, since a real part for $\tau$ would complicate the duality.}. One must change all pairs of $(n^i,m^i)$ along this section by $(m^i,n^i)$. 
In this image are the $m$'s who play the role of magnetic flux quanta. Also, a D3 in one image is mapped to a D9 in the other image, and a D7 in one image goes to a D5 in the other one.
Finally, we would like to recall that one can generate lower dimensional D-brane charges in the world-volume of higher dimensional ones when turning on magnetic fluxes on
them, or, alternatively, any D$p$-brane can be seen as a D$p'$-brane, $p<p'$, with magnetic flux on it. Thus, a D5-brane can be seen as a D9-brane with infinite magnetic flux in one torus
and minus infinite magnetic flux in another torus. This implies, as emphasized in the main text, that, at least in the absence of RR and NSNS fluxes, one can see every previously considered 
toroidal inflationary model as coming just from Type I D9 branes with magnetic fluxes and Wilson lines turned on.

\section{A concrete model}\label{app_model}

As already mentioned, Wilson line inflation provides a T-dual image of 
angled-brane inflation (studied in detail in \cite{rabadan})
and, as such, its predictions are rather similar to this form of brane 
inflation in a large set of situations, albeit with some subtleties like 
the role played in this situation by
the momentum modes. However, we want to stress that there are some situations in which Wilson lines and brane positions are not entirely equivalent in a given physical situation (though,
of course, T-duality will change the roles of both kinds of fields). 
There are some situations, like compactifications with RR and NSNS fluxes, 
that are
highly asymmetric with respect to Wilson lines and brane positions: for example, brane positions get generically a (either soft or supersymmetric) mass from the effect of these fluxes whereas
the potential for Wilson lines remains flat. We want to explore this fact 
to get inflation with Wilson lines in a situation in which the relative 
positions between two stacks of branes
are stabilised.

In the following subsection we will review the mechanism (discovered in \cite{soft} and reinterpreted in an elegant way in \cite{landscape}; see also \cite{martucci} for closely related work) by which brane positions are stabilised in
the presence of supersymmetric $G_3$ flux, and show how and where brane positions are fixed. We present this model to show how our model can be implemented in a fluxed Type IIB setup.

\subsection{Fixing D7 brane positions with fluxes}
Consider Type II string theory compactified on the $T^6/Z_2$ orientifold considered by \cite{kst}. Consider a RR and NSNS flux of the form
\beqa
{1\over 4\pi^2\a'}F_3&=&a_0 \a_0 + a\sum_i \a_i + b\sum_i \b_i + b_0 \b_0\label{F3}\\
{1\over 4\pi^2\a'}H_3&=&c_0 \a_0 + c\sum_i \a_i + d\sum_i \b_i + d_0 \b_0\label{H3} ,
\eeqa
where the $a,b,c,d$ and the $a_0$'s etc are integers and the $\a_a$'s and 
$\b_a$'s form a basis of real 3-forms:
\beqa
\a_0&=&dx^1\w dx^2 \w dx^3\nonumber\\
\b_0&=&dy^1\w dy^2 \w dy^3\nonumber\\
\a_i&=&dy^i\w dx^j \w dx^k\\
\b_i&=&-dx^i\w dy^j \w dy^k\nonumber , 
\eeqa
where the order of the indices in $\a_i$ and $\b_i$ is defined to be the canonical one. We are going to use the mechanism described in \cite{landscape} to fix the moduli of the branes at
different positions in the torus. Consider a D7 brane wrapping the second and third torus, and being point-like in the first one. Given the $H_3$ flux (\ref{H3}), we can find a
local description of $B_2$ in a patch that is convenient for us:
\beqa
B_2&=&4\pi^2\a'\{(x^1c_0+y^1c)dx^2\w dx^3+(-x^1d+y^1d_0)dy^2\w dy^3\nonumber\\
&&+(x^1c-y^1d)(dx^2\w dy^3+dy^2\w dx^3)\} . 
\eeqa
Consider the following magnetic flux in the D7-brane\fn{We are using the symbols $F_2$ for the magnetic field and $F_3$ for the RR field. We hope this rather conventional choice will not 
confuse the reader.}
\beqa
F_2=2\pi\left(\a~ dx^2\w dx^3+ \b~ dy^2\w dy^3+\g~ (dx^2\w dy^3 + dy^2\w dx^3)\right)
\eeqa
with $\a$, $\b$, $\g$ integers. The supersymmetry condition for the D7 implies 
$B|_{D7}+2\pi \a' F_2=0$, which is translated into the equations\fn{Note that 
satisfaction of these equations implies that no tadpole cancellation is  
required for the numbers $\a$, $\b$, $\g$.}
\beqa
x_1c_0+y_1c+\a&=&0\nonumber\\
-x_1d+y_1d_0+\b&=&0\label{eqns}\\
x_1c-y_1d+\g&=&0\nonumber . 
\eeqa
Since the only variables here are $x_1,y_1$, the system is overdetermined. One must stress however that the flux integers $c,d$'s and the magnetic integers $\a$, $\b$, $\g$ are not
free parameters but are subject to some other conditions. At this point, it is better to take a precise example from where we extract the flux parameters and see what are the conditions on
$F$. Before doing that, we can compute the mass corresponding to the open string field whose vev gives the position of the D7. It is given by \cite{landscape, soft}
\beqa
m^2={g_s\over 2}|G_{\bar{1}23}|^2 ,
\eeqa
where $G_3=F_3-\phi H_3$, with $\phi=C_0+i/g_s$ the axio-dilaton. 
Now, from (\ref{F3}), (\ref{H3}) we get, using $dz^i=dx^i+\tau^i dy^i$,
\beqa
F_{\bar{1}23}=K(\tau^i)\left( a_0~\tau^1\bar\tau^2\bar\tau^3-a~(\tau^1\bar\tau^2+\tau^1\bar\tau^3+\bar\tau^2\bar\tau^3)-
b(\tau^1+\bar\tau^2+\bar\tau^3)-b_0\right)\\
H_{\bar{1}23}=K(\tau^i)\left( c_0~\tau^1\bar\tau^2\bar\tau^3-c~(\tau^1\bar\tau^2+\tau^1\bar\tau^3+\bar\tau^2\bar\tau^3)-
d(\tau^1+\bar\tau^2+\bar\tau^3)-d_0\right)
\eeqa
with
\beqa
K(\tau^i)\equiv {i\pi^2 \a'\over 2\prod_i\pim \tau^i (2\pi)^3 R_1R_2R_3} .
\eeqa
Note that the prefactor $K$ encodes all the dependence of the masses in the  
K\"ahler moduli. Let us take for example the values\fn{To avoid the kind of  
subtleties pointed out in \cite{kst}, \cite{n3}, the values of $n_1$,  
$n_2$ can be restricted to be even.}
$(a_0,a,b,b_0)=n_1(1,0,0,1)$, $(c_0,c,d,d_0)=n_2(1,-1,-1,-2)$ from \cite{kst}, with $n_i$ integers. These integers fix the complex structure moduli and dilaton to the values
\beqa
\tau^i=e^{2\pi i\over 3}\equiv \tau, \ \, \quad \phi={n_1\over n_2}e^{2\pi i\over 3}={n_1\over n_2}\tau.
\eeqa
This implies $g_s=(2 n_2/\sqrt{3}n_1)$. These fluxes produce a D3 tadpole
\beqa
N_{flux}={1\over (2\pi)^4\a'^2}\int H_3\w F_3=2n_1n_2.
\eeqa
For these values, the equations (\ref{eqns}) become
\beqa
x_1-y_1&=&-\a/ n_2\nonumber\\
x_1-2y_1&=&-\b/ n_2\label{eqns2}\\
-x_1+y_1&=&-\g/ n_2.\nonumber
\eeqa
One possible solution is to take $\a=-\g$. Then the position of the brane
is fixed at
\beqa
x_1={\b-2\a\over n_2}, \ \quad y_1={\b-\a\over n_2}.
\label{positions}
\eeqa
We can also compute the mass of the open string excitation corresponding to the position of the brane. The relevant component of the $G_3$ flux is given by
\beqa
G_{\bar{1}23}=-n_1K(\tau)(\tau+2)^2
\eeqa
So
\beqa
m^2={n_1^2g_s\over 512 \pi^2 (\pim \tau)^7}(5+4~\preal\tau)^2{\a'^2\over (R_1R_2R_3)^2}={n_1n_2\over 12\sqrt{3}\pi^2}{\a'^2\over (R_1R_2R_3)^2}.\label{mass}
\eeqa
We must emphasize that this mass term is supersymmetric. On the other hand, there is no such constraint for Wilson lines. As emphasized in \cite{soft}, this is due to the fact that
generic stabilisation of Wilson lines corresponding to a given D7 brane would be incompatible with gauge invariance.

We are computing the inflationary potential assuming that the positions are stabilised, but, in a given situation, we must check that this is indeed the case. To see this, consider a situation
in which the Wilson lines are put to zero and the position field is expressed in terms of some canonically normalised field $\phi=y/2\pi \a'$, $y$ being the position of a brane \cite{soft}. 
The mass (\ref{mass}) is computed for this field. The potential for $y$ near
the minimum $\phi_0=2\pi\a'y_0$, with $y_0$ given in (\ref{positions}) is roughly
\beqa V(\phi)\simeq m^2(\phi-\phi_0)^2-{k\over \phi^2}
\eeqa
with $k$ arbitrarily small when $\theta\to 0$.
The addition of the $\phi^{-2}$ part to the quadratic potential will only manifest itself in scales of order $k\over \phi_0^4$ around $\phi_0$. That means that we can consider the field $\phi$
to be stabilised whenever
\beqa
m^2\gg{k\over\phi_0^4}.
\eeqa
that can be satisfied easily just making the angle small. This basically means that any attempt of getting inflation with some field related to the position of a D7-brane in the presence 
of $G_3$ flux will necessarily face the issue that this field will unavoidably 
receive a non-negligible positive contribution to its mass coming from its 
backreaction to the flux.

%\newpage

\end{document}